# The influence of the inverse Compton effect on the transverse momentum spectra of particles produced in *pp* collisions at $\sqrt{s} = 14$ TeV.

M. Alizada and M. Suleymanov
Baku State University

## Abstract

The influence of the QED-analog of the inverse Compton effect on the transverse momentum spectra of particles produced in proton-proton collisions at energies of $\sqrt{s} = 14\,TeV$ has been investigated. The analysis is based on the quark-gluon scattering process $g + q \rightarrow g + q$, which is the QCD analogue of Compton scattering of a photon on an electron and can lead to energy redistribution between partons, analogous to the mechanism of the inverse Compton effect.

Data obtained numerically using the **PYTHIA** event generator (version 8.316) were used. A total of *5×10⁵* proton-proton collisions at *√s = 14 TeV* were analyzed. Events were classified based on the relative energies of the initial quark and gluon, which allowed us to distinguish Compton scattering (DCE) events from inverse Compton scattering (ICE) events. Particle transverse momentum spectra were obtained in the region: *$p_T$ <10 GeV/c*.

The results showed that including inverse Compton scattering events in the analysis leads to a moderate increase in particle yield. The ratio of the spectra for ICE and DCE events remains approximately constant and is about 1.1 within statistical errors. No significant broadening of the transverse momentum spectra is observed. These results show that proton-proton collisions can serve as a reliable baseline for studies of energy redistribution mechanisms in a dense QCD medium, such as quark-gluon plasma.

## 1. Introduction

The origin of ultra-high-energy cosmic rays (UHECR, $E \geq 10^{18}$ eV) [1] remains one of the central problems of modern astrophysics and high-energy physics. The observed high-energy portion of the cosmic ray spectrum, $E \geq 10^{18}$ eV, cannot be explained by known galactic particle acceleration mechanisms; therefore, the existence of extragalactic sources for these particles is postulated. The book [2] discusses the limits of galactic accelerators and the necessity of extragalactic sources for energies above $\sim 10^{17}$–$10^{18}$ eV. The review [3] asserts that galactic sources cannot explain the spectrum above the "ankle" ($\sim 10^{18}$ eV), and an extragalactic contribution is required, while the paper [4] discusses observations and the interpretation of UHECRs as extragalactic particles.

In [5–7], a mechanism for the production of ultra-high-energy particles was proposed, based on the acceleration of partons in a dense, strongly interacting nuclear medium. In terms of its structure, this mechanism is analogous to inverse Compton scattering, known in quantum electrodynamics (QED), where a relativistic electron transfers energy to a photon. In the case under consideration, energy redistribution occurs within the quark-gluon medium between parton degrees of freedom.

Under astrophysical conditions, this mechanism may be further amplified by multiple interactions of particles with a heterogeneous (turbulent) medium, leading to a stochastic increase in their energy characteristic of second-order Fermi acceleration, in which particles interact multiple times with moving magnetohydrodynamic inhomogeneities of the medium (see, e.g., [8]). In the simplest model, the energy of a particle after $N$ interaction acts are described by the relation

$$E_N \approx E_0(1+\delta)^N,$$
which, in the continuous limit, leads to an exponential growth law
$$E \sim E_0 e^{\delta N}.$$
Taking into account energy losses due to radiative and interaction processes, the energy evolution is described by the following equation:
$$\frac{dE}{dN} = \delta E - \beta E,$$
where the parameter $\delta$ characterizes the acceleration efficiency, and $\beta$ represents the total energy losses. Under the condition $\delta > \beta$, an effective acceleration regime analogous to the Fermi mechanism is realized, making this model relevant for describing the generation of high-energy particles in astrophysical sources.

In quantum chromodynamics (QCD), there are processes that can be regarded as QCD analogues of the Compton scattering of a photon by an electron in QED. This reaction, described by quark-gluon scattering
$$g + q \to g + q.$$
Reference [9] shows that such processes lead to the production of gluon jets in photon–hadron collisions and direct photons in hadron–hadron collisions. Cross sections are calculated using quark and gluon hadron distribution functions, which naturally allows for energy redistribution among parton degrees of freedom.

At high energy, the dominant contribution comes from momentum transfer in the t-channel, where:
$$t = (p_1 - p_3)^2 \simeq -p_T^2$$
Small $|t|$ (i.e., low $p_t$) enhances the amplitude, resulting in a characteristic t-channel enhancement.

Let us consider the kinematics of a 2→2 parton process:
$$a(p_1) + b(p_2) \to c(p_3) + d(p_4)$$
where *a, b, c, d* are quarks or gluons.

Feynman diagrams for the $gq \to gq$ process were drawn using the FeynArts program [10] (Fig. 1(a,b,c)).

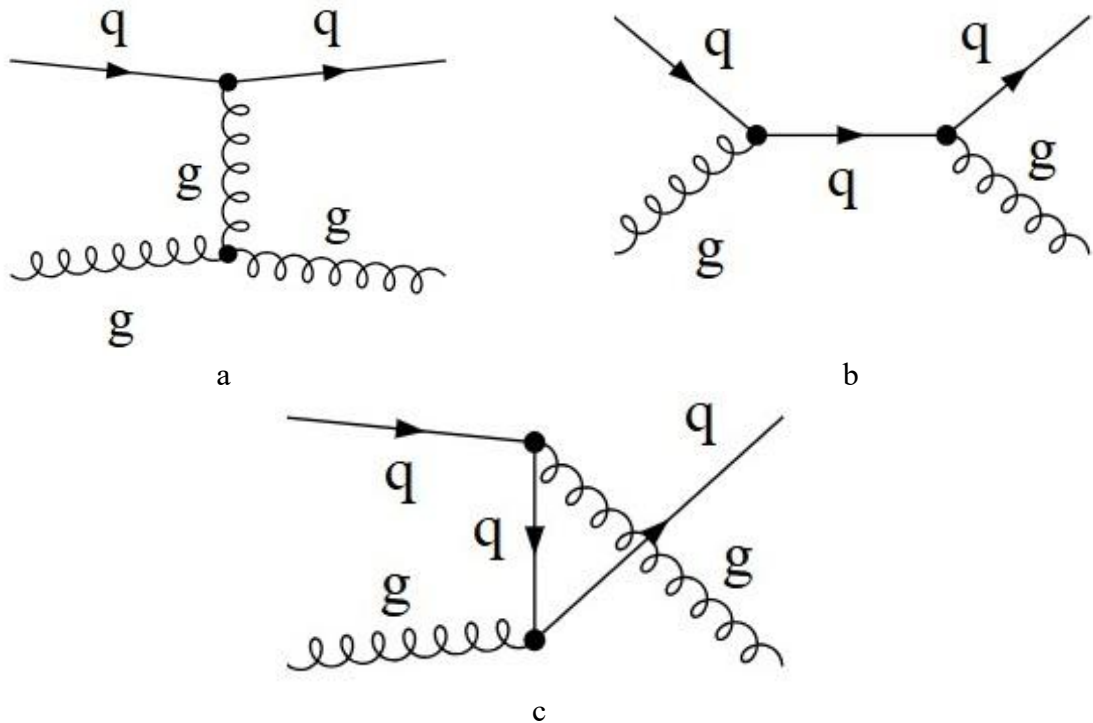

Figure 1. Feynman diagrams for the process $gq \to gq$.

Mandelstam variables:
$$\hat{s} = (p_1 + p_2)^2; \quad \hat{t} = (p_1 - p_3)^2; \quad \hat{u} = (p_1 - p_4)^2$$
and for massless partons
$$\hat{s} + \hat{t} + \hat{u} = 0$$
In hadron collisions $\hat{s} = x_1 x_2 s$, where $x_1 x_2$ are the parton momentum fractions.

Process matrix elements $gq \to gq$

$$M_1 = -\frac{g_s^2 T^{Glu2}_{Col1 Col5} T^{Glu4}_{Col3 Col5} \bar{U}(p_2) \hat{\varepsilon}^*(k_2)(\hat{k}_2 + \hat{p}_2 + m_q)\hat{\varepsilon}(k_1) U(p_1)}{(-k_2 - p_2)^2 - m_q^2}$$

$$M_2 = \frac{2ig_s^2 T^{Glu5}_{Col3 Col1} f^{Glu2\,Glu4\,Glu5} \hat{k}_1 \hat{\varepsilon}^*(k_2) \bar{U}(p_2)\hat{\varepsilon}(k_1) U(p_1)}{(k_2 - k_1)^2} +$$
$$\frac{2ig_s^2 T^{Glu5}_{Col3 Col1} f^{Glu2\,Glu4\,Glu5} \hat{k}_2 \hat{\varepsilon}(k_1) \bar{U}(p_2)\hat{\varepsilon}^*(k_2) U(p_1)}{(k_2 - k_1)^2} +$$
$$\frac{ig_s^2 T^{Glu5}_{Col3 Col1} f^{Glu2\,Glu4\,Glu5} \hat{\varepsilon}(k_1) \hat{\varepsilon}^*(k_2) \bar{U}(p_2)(-\hat{k}_1 - \hat{k}_2) U(p_1)}{(k_2 - k_1)^2},$$

$$M_3 = -\frac{g_s^2 T^{Glu4}_{Col5\,Col1} T^{Glu2}_{Col3\,Col5} \bar{U}(p_2) \hat{\varepsilon}(k_1)(\hat{p}_2 - \hat{k}_1 + m_q) \hat{\varepsilon}^*(k_2) U(p_1)}{(k_1 - p_2)^2 - m_u^2}.$$

Differential cross section:
$$\frac{d\hat{\sigma}(gq \to gq)}{d\hat{t}} = \frac{1}{16\pi\hat{s}^2} |\bar{M}|^2.$$

For the square of the process matrix element $gq \to gq$, averaged over the spins of the final particles and calculated using FeynCalc [11], we obtain:
$$\overline{|M|^2} = -\frac{g_s^4(s^2 + u^2)(9s^2 - t^2 + 9u^2)}{18 st^2 u}$$

At low $|\hat{t}^2|$, t-channel production leads to sharper $p_T$ spectra.

In the center-of-mass system, the energy of the final parton
$$E_3^* = \frac{\sqrt{\hat{s}}}{2}$$

In the laboratory system
$$E_3 = \gamma(E_3^* + \beta p_3^* \cos\theta^*).$$

If a fast parton with energy $E_{fast}$ interacts with a medium parton with energy $E_{med}$,
$$\hat{s} \approx 2 E_{fast} E_{med} (1 - \cos\theta)$$

Structurally, this is analogous to the inverse Compton effect:
$$E' \sim \gamma^2 E.$$

Gain
$$K = \frac{E_{out}}{E_{in}} \sim \gamma_{eff}^2 \frac{E_{med}}{E_{fast}},$$

Thus, a dense quark-gluon medium can redistribute energy, leading to the appearance of particles with elevated $p_T$. It was expected that such a superdense medium could form in heavy-ion collisions at colliders: SPS CERN [12], RHIC BNL [13]; LHC CERN [14]. Experimental data obtained from these colliders confirm the formation of a dense nuclear medium with collective hydrodynamic properties [15]. In nature, such highly compressed states of matter can occur in neutron stars (at the center of a star) [16] (Neutron star), in quark stars (Quark star) [17] as a result of the collapse of supernovae, as well as in the early Universe during the "Big Bang" [18] and during hypothetical "mini Big Bangs."

## 2. Methodology

In the introduction, we noted that a dense quark-gluon medium can redistribute parton energy, leading to the appearance of particles with elevated $p_T$ as a result of ICE during the process $gq \to$

$gq$. That is, studies of $p_T$ distributions in hadron-hadron, hadron-nucleus, and nucleus-nucleus collisions at the high energies achieved at RHIC and LHC may provide observational evidence for the existence of the parton acceleration effect in ICE during the process $gq \to gq$. The main objective of this work is to investigate the influence of ICE in the process $gq \to gq$ on the $p_T$ spectra of particles produced in pp collisions at the maximum total energy $\sqrt{s} = 14$ ТэВ (LHC CERN).

To obtain these spectra, the **PYTHIA 8.316** event generator [19] was used in this work. The simulation included the main hard parton–parton interaction processes of leading order in the QCD theory, namely:
- glueon-glueon scattering, the dominant process at LHC energies;
- gluon annihilation with quark–antiquark pair production;
- quark-gluon scattering;
- scattering of quarks of different flavors;
- scattering of quarks of the same flavor;
- quark–antiquark annihilation with the creation of a pair of a different flavor;
- quark–antiquark annihilation into gluons;
- elastic quark–antiquark scattering.

In accordance with the above, in the KHD code under consideration, the analogue of Compton scattering is the quark-gluon scattering processes $gq \to gq$.

In these processes, a fast parton (quark or gluon) scatters off a non-quasi-classical gluon field (in pp collisions), energy and momentum are redistributed, and one of the outgoing partons may acquire a significantly large $p_T$.

The **PYTHIA 8.316** event generator was used for the numerical simulation of the process under consideration. A total of 500,000 *pp* events (at energies of 14 TeV) were simulated without accounting for the inverse Compton effect (designated DCE, i.e., when the gluon is faster than the quark) and 500,000 *pp* events accounting for the inverse Compton effect (designated ICE, i.e., when the quark is faster than the gluon).

To isolate the net contribution of QHD-Compton scattering, other channels of hard QHD interactions, such as gluon-gluon ($gg \to gg$) and quark-quark ($qq \to qq$) scattering, were forcibly disabled. Only the quark-gluon scattering process $qg \to qg$ remained active.

To describe the internal structure of colliding protons, the **CT14QED** set of parton distribution functions (PDFs) was used in the simulation.

Since no explicit restrictions on the parton shower stage were applied in the generator configuration, radiation effects in the initial and final states (ISR and FSR) were enabled by default. This allows for the consideration of multiple gluon emissions and the realistic simulation of parton evolution before and after the hard scattering event.

In addition, the hadronization stage, implemented in Pythia 8.316 based on the Lund string fragmentation model, was included in the simulation. The inclusion of the for hadronization and parton showers is fundamental to this study, as it allows us to describe the transition from colored partons to the observed colorless final-state particles, for which the kinematic distributions were subsequently constructed.

The algorithm for analyzing each generated event (pythia.next()) included the following steps:
1. **Identification of hard partons** participating in the primary scattering.
2. **Classification of the event** as "direct" or "reverse" based on a pairwise comparison of the energies of the selected initial partons.
3. **Selection of final-state hadrons**. From the entire set of particles, only stable final hadrons were selected using the built-in methods of the Particle class:
   if (!p.isFinal() || !p.isHadron()) continue;
4. **Filling histograms**. For each selected hadron, $p_T$ and $\eta$ were calculated, after which one-dimensional distributions over $p_T$ and two-dimensional distributions $(\eta, p_T)$ were filled, separated by event type.

As a result of the algorithm's operation, distributions were generated for transverse momentum in the range $p_T \in [0,10]$ GeV/c and pseudorapidity $\eta \in [-10,10]$. To obtain correct probability distributions (particle densities in phase space), all final histograms were normalized by the number of events in the corresponding class.

## Results.

Based on simulations performed using the method described above, distributions of secondary particles over transverse momentum were obtained. Figure 2 shows the inclusive cross sections for the production of secondary particles $\frac{d\sigma}{dp_T}$ in proton-proton collisions at energies $\sqrt{s} = 14$ ТэВ.

The white dots correspond to cases of Compton scattering (DCE), while the black dots describe events in which the inverse Compton mechanism (ICE) is realized. For a quantitative analysis of the spectral behavior, the ratio

$$Ratio = \frac{\left(\frac{d\sigma}{dp_T}\right)_{ICE}}{\left(\frac{d\sigma}{dp_T}\right)_{DCE}}$$

The results show that, within statistical errors, the value of $Ratio \cong 1.1$ is practically independent of $p_T$. This indicates a systematically higher normalization of inclusive cross sections in ICE events compared to DCE events.

To explain this result, recall that for processes of the type $gq \rightarrow gq$, as noted above, the cross section behaves as follows

$$\frac{d\sigma}{dp_T^2} \sim \frac{\alpha_s^2}{p_T^4}$$

at low $p_T$ ($\alpha_s$ is the strong interaction constant in QCD, in 2→2 scattering processes $M \sim \alpha_s$, and $\sigma \sim M^2 \sim \alpha_s^2$); ICE makes more effective use of the gluon in the initial state as a "hard" parton. In ICE, the contribution from t-channel gluon exchange is stronger, which automatically enhances the cross section at low and intermediate $p_T$ and results in a slower tail decay. This was the effect of kinematics. Another important factor influencing the behavior of inclusive cross sections as a function of $p_T$ is the structure of the parton distribution within the hadron (PDF effect).

For a fixed $p_T$, one can simply write that:

$$p_T \sim \sqrt{x_1 x_2 s}$$

ICE is more likely to occur at lower $x_{(g)}$ values. At energies of 14 TeV: $g(x) \gg q(x)$ when $x \lesssim 10^{-2}$. Therefore, for each bin of $p_{(T)}$, the probability of ICE is higher simply due to the PDF. In the region $p_T > 5\ GeV$, two factors come into play simultaneously: the color factors $C_F = \frac{4}{3}$ for the quark and $C_A = 3$ for the gluon; therefore, in ICE, the values of ISR (initial state radiation) are larger than in the case of DCE, since ISR is proportional to the color factor and the values of $C_A > C_F$, the gluon emits more frequently. This means that

$$P_{emission} \propto \alpha_s C_i$$

($C_i$ is the color factor (Casimir factor), which depends on the type of the emitting particle). That is, for a gluon, the probability of radiation is almost twice as high.

Thus, the results obtained indicate that the inclusion of ICE does not lead to a significant hardening of the transverse momentum spectrum (i.e., a shift of the particle distribution in $p_T$ toward higher transverse momenta), but only to a moderate increase in normalization.

## Conclusion

An analysis of the quark-gluon scattering process $gq \rightarrow gq$ in proton-proton collisions at energies of $\sqrt{s} = 14$ ТэВ was performed to investigate the effect of the KHD analogue of the inverse Compton effect on the transverse momentum spectra of secondary particles.

The results show that in pp collisions, this process forms the standard kinematic structure *of 2→2-type* parton scattering within the leading order of quantum chromodynamics. The behavior of the spectra is determined primarily by t-channel enhancement and the structure of parton distribution functions in hadrons.

Including the ICE mechanism does not lead to a noticeable broadening of the transverse momentum spectra, but it does cause a moderate increase in the overall normalization of the inclusive cross sections. Thus, pp collisions can be considered a reliable baseline for subsequent analysis of dense medium effects in heavy-ion collisions.

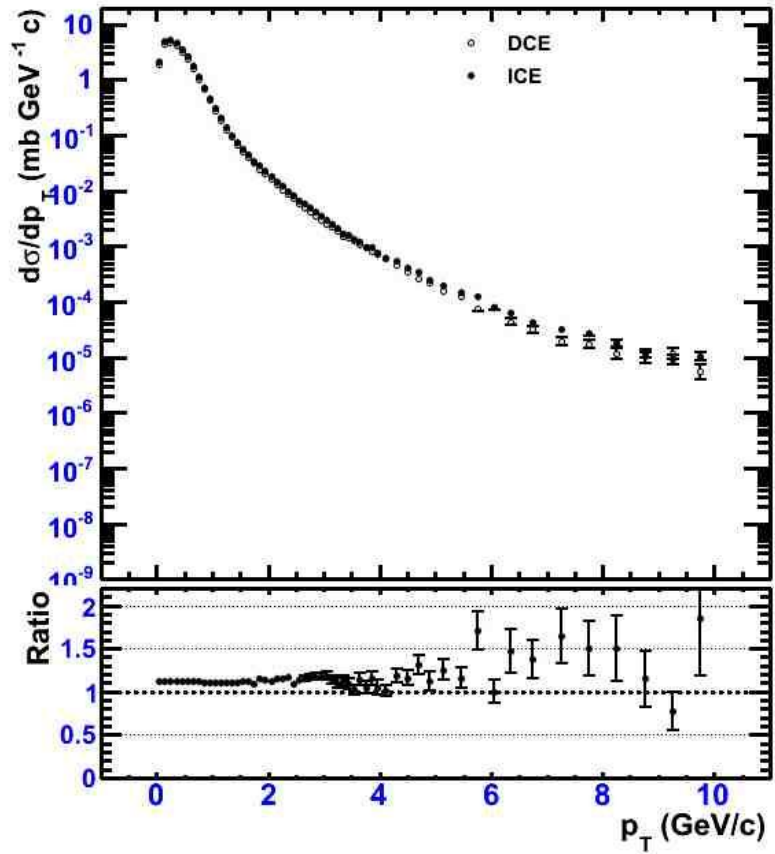

**Figure 2**. Inclusive cross sections for the production of secondary particles (top panel) formed in 14 TeV pp collisions. The ratio of inclusive cross sections in ICE events to those for particles produced in DCE events (bottom panel)

**References**
[1] A.V. Olinto. Ultra-high-energy cosmic rays: the theoretical challenge. Physics Reports Volumes 333–334, August 2000, Pages 329–348]
[2] V. S. Berezinsky et al Astrophysics of Cosmic Rays, North-Holland, 1990 - Science - 534 pages
[3] Physics and Astrophysics of Ultra High Energy Cosmic Rays, Book, 2001 1st edition. Editors: Martin Lemoine and Günter Sigl
[4] A. A. Watson, "High-energy cosmic rays and the Greisen–Zatsepin–Kuzmin effect," Reports on Progress in Physics, Volume 77, Number 3
[5] Suleymanov M 2016 Georgian Electronic Scientific Journal: Physics 1(15) 92
[6] Suleymanov M 2009 Proceedings of Science EPS-HEP2009 406
[7] Suleymanov M 2012 J. Phys.: Conf. Ser. 347 012024
[8] E. Fermi. On the Origin of Cosmic Radiation. Phys. Rev. 75, 1169 – Published April 15, 1949
[9] Fritzsch H and Minkowski P 1977 Phys. Lett. B 69 316–320
[10] Hahn T 2001 Comput. Phys. Commun. 140 418–431 (FeynArts)
[11] Shtabovenko V, Mertig R, Orellana F 2025 Comput. Phys. Commun. 306
[12] https://home.cern/science/accelerators/super-proton-synchrotron 87
[13] https://www.bnl.gov/rhic/ 12,13  87, 98

[14] https://home.cern/science/accelerators/large-hadron-collider 109–1312 .
[15] Ulrich Heinz and Raimond Snellings. "Collective Flow and Viscosity in Relativistic Heavy-Ion Collisions." *Annual Review of Nuclear and Particle Science*, 65, 1312.
[16] Norman K. Glendenning. Compact Stars: Nuclear Physics, Particle Physics, and General Relativity. Textbook. 1997. 1st edition, pp. 1413, 151
[17] Edward Witten·Cosmic separation of phases. Phys. Rev. D 30, 272 – Published July 15, 1984 DOI: https://doi.org/10.1103/PhysRevD.30.272 1413–1918
[18] Steven Weinberg. The First Three Minutes. A Modern View of the Origin of the Universe. FLAMINGO Published by Fontana Paperbac. 1977, 2165, 2276
[19] Sjöstrand T et al. 2015 Comput. Phys. Commun. 191 159–177